\documentclass[aip,amsmath,amssymb,reprint,floatfix]{revtex4-1}
\usepackage[T1]{fontenc}
\usepackage[utf8]{inputenc}
\usepackage{color}
\definecolor{page_backgroundcolor}{rgb}{1, 1, 1}
\pagecolor{page_backgroundcolor}
\usepackage{array}
\usepackage{multirow}
\usepackage{amsmath}
\usepackage{graphicx}
\usepackage{esint}

\makeatletter

\providecommand{\tabularnewline}{\\}

\usepackage{array}
\usepackage{enumitem}
\usepackage{afterpage}
\usepackage{caption}
\usepackage{multirow}

\makeatother

\begin{document}
\title{Real Space Orthogonal Projector-Augmented-Wave Method}
\author{Wenfei Li and Daniel Neuhauser}
\begin{abstract}
The projector augmented wave (PAW) method of Blöchl makes smooth but
non-orthogonal orbitals. Here we show how to make PAW orthogonal,
using a cheap transformation of the wave-functions. We show that the
resulting Orthogonal PAW (OPAW), applied for DFT, reproduces (for
a large variety of solids) band gaps from the ABINIT package. OPAW
combines the underlying orthogonality of norm-conserving pseudopotentials
with the large grid spacings and small energy cutoffs in PAW. The
OPAW framework can also be combined with other electronic structure
theory methods.
\end{abstract}
\maketitle

\section{Introduction}

A plane wave basis set is natural when studying periodic systems with
DFT and post-DFT methods. Convergence with basis set is simply verified
by increasing a single parameter, the kinetic energy cutoff. However,
due to the fast oscillation of atomic core states, a direct all-electron
treatment is prohibitive. One way to circumvent this problem is to
replace the effect of the chemically inert core states by an effective
pseudo-potential, and the resulting pseudo valence states are non-oscillatory.\citet{reis2003first,willand2013norm}
DFT using pseudo-potentials and a plane wave basis set has therefore
become one of the most popular choices in computational chemistry
and materials science. However, despite the formal simplicity of norm-conserving
pseudo-potentials (NCPP), treatment of first-row elements and transition
metals is still computationally demanding, due to the localized nature
of $2p$ and $3d$ orbitals.\citet{kresse1994norm-conserving,hamann1979norm,hamann2013optimized}

The projector-augmented wave (PAW) method proposed by Blöchl\citet{blochl1994projector,kresse1999ultrasoft,blochl2003projector,holzwarth1997comparison}
seeks to make softer pseudo wavefunctions by relaxing the norm-conserving
condition. There are several different implementations of the PAW
method (e.g., \citet{tackett2001projector,torrent2008implementation,enkovaara2010electronic,mortensen2005real})
with many successful applications.

In addition to the reduced kinetic energy cutoff, an advantage of
the PAW method is that it provides means for recovering the all-electron
orbitals, and these orbitals possess the right nodal structures in
the core region. Therefore, PAW enables the calculation of quantities
such as hyperfine parameters, core-level spectra, electric-field gradients,
and the NMR chemical shifts, which rely on a correct description of
all-electron wavefunctions in the core region.\citet{pickard2001all-electron}

The PAW method is based on a map between the smoothed pseudo wavefunctions
$\{\tilde{\psi}_{m}\}$ and the all electron wavefunctions $\{\psi_{m}\}$.
Unlike NCPP where the wavefunctions retain their orthogonality, the
pseudo wavefunctions in PAW satisfy a generalized orthogonality condition:
\begin{equation}
\langle\tilde{\psi}_{m}|\hat{S}|\tilde{\psi}_{n}\rangle=\delta_{mn},\label{eq:psitilde_S_psitilde}
\end{equation}
which leads to a generalized eigenproblem: $\tilde{H}\tilde{\psi}_{m}=\epsilon_{m}\hat{S}\tilde{\psi}_{m}$
where we introduced the 1-body Hamiltonian $\tilde{H}$ and overlap
operator $\hat{S}$ (both detailed later).

The fact that the pseudo-orbitals are not orthogonal complicates,
however, the use of PAW for applications that rely on the orthogonality
of molecular orbitals. These include some post-DFT methods, as well
as several lower-scaling DFT methods, including the modified deterministic
Chebyshev approach (see, e.g., \citet{zhou2014chebyshev-filtered})
or stochastic DFT methods,\citet{baer2013self,EmbeddedFragments2014}
which are able to handle a large number of electrons (potentially
hundreds of thousands for the stochastic approach) by filtering a
function of an orthogonal Hamiltonian.

Here we solve the non-orthogonality problem by an efficient numerical
transformation of the PAW problem to an orthogonal one, 
\begin{equation}
\left(\hat{S}^{-\frac{1}{2}}\tilde{H}\hat{S}^{-\frac{1}{2}}\right)\bar{\psi}_{m}=\epsilon_{m}\bar{\psi}_{m}\label{eq:HrotPsi_Epsi}
\end{equation}
with $\bar{\psi}_{m}=\hat{S}^{1/2}\tilde{\psi}_{m}$ forming an orthogonal
set, with the same norm as the all-electron orbitals (to be proved
later). The key is that we show how to numerically apply the $\hat{S}^{-1/2}$
(or $\hat{S}^{-1})$ operator efficiently, without significantly raising
the cost of applying the Hamiltonian.

The resulting approach retains one of the desirable features of NCPP,
orthogonality of molecular orbitals, and we therefore label it Orthogonal
PAW (OPAW). In addition to orthogonality, OPAW is also efficient because
it is implemented in real space, exploiting the localization of atomic
projector functions and partial waves.\citet{enkovaara2010electronic,mortensen2005real}

OPAW provides a general framework, and can be combined with different
electronic structure methods. Here we apply the method with the Chebyshev-filtered
subspace iteration (CheFS) DFT approach, concentrating on the fundamental
band gap of solids. We show below excellent agreement with PAW calculations
from the ABINIT package.\citet{torrent2008implementation,Gonze2020}
We also demonstrate that for many systems, PAW and OPAW band gaps
converge with energy cutoff faster than NCPP.

Section II presents the OPAW theory. Results are presented in Section
III, and conclusions follow in Section IV. Technical details are deferred
to appendices.

\section{Theory}

\subsection{Orthogonal projector augmented wave}

The basic relation in PAW is a map $\hat{T}$ yielding the true molecular
eigenstates, $\psi_{m}$, from the smoother pseudo-orbitals 
\begin{equation}
|\psi_{m}\rangle=\hat{T}|\tilde{\psi}_{m}\rangle\equiv|\tilde{\psi}_{m}\rangle+\sum_{a,i}\left(|\phi_{i}^{(a)}\rangle-|\tilde{\phi}_{i}^{(a)}\rangle\right)\langle p_{i}^{(a)}|\tilde{\psi}_{m}\rangle,
\end{equation}
where $a$ is the atom index and $i$ runs over all the partial wave
channels (a combination of principal, angular momentum and magnetic
quantum numbers) associated with each atom; $\phi_{i}^{(a)}$ and
$\tilde{\phi}_{i}^{(a)}$ are a true atomic orbital and a smoothed
version which matches $\phi_{i}^{(a)}$ outside a small sphere around
the atom (labeled the augmentation region). The atomic projectors
$\{p_{i}^{(a)}\}$ are localized in the augmentation region, and are
built to span the space within each augmentation sphere, i.e., $\sum_{i}|\tilde{\phi}_{i}^{(a)}\rangle\langle p_{i}^{(a)}|\simeq1$
in the sphere.

With some derivations, one arrives at the working equation of PAW,
the generalized eigenproblem $\tilde{H}\tilde{\psi}_{m}=\epsilon_{m}\hat{S}\tilde{\psi}_{m}$
where 
\begin{equation}
\hat{S}\equiv\hat{T}^{\dagger}T=\mathbb{I}+\sum_{ij,a}|p_{i}^{(a)}\rangle s_{ij}^{(a)}\langle p_{j}^{(a)}|,\label{eq:sij}
\end{equation}
with $s_{ij}^{(a)}\equiv\left\langle \phi_{i}^{(a)}|\phi_{j}^{(a)}\right\rangle -\left\langle \tilde{\phi}_{i}^{(a)}|\tilde{\phi}_{j}^{(a)}\right\rangle ,$
and 
\begin{equation}
\tilde{H}=-\frac{\nabla^{2}}{2}+\nu_{KS}(\boldsymbol{r})+\sum_{ij,a}|p_{i}^{(a)}\rangle D_{ij}^{(a)}\langle p_{j}^{(a)}|.\label{eq:H}
\end{equation}

The expressions for the Kohn-Sham effective potential $\nu_{KS}(\boldsymbol{r})$
and for $D_{ij}^{(a)}$ are found in various references.\citet{blochl1994projector,torrent2008implementation}
While $s_{ij}^{(a)}$ are only atom-dependent, $\nu_{KS}(\boldsymbol{r})$
and $D_{ij}^{(a)}$ both depend on the on-site PAW atomic density
matrices: $\rho_{ij}^{(a)}=\sum_{m}\langle p_{j}^{(a)}|\tilde{\psi}_{m}\rangle\langle\tilde{\psi}_{m}|p_{i}^{(a)}\rangle$,
as well as the smooth density $\tilde{n}(\boldsymbol{r})=\sum_{m}\left|\tilde{\psi}_{m}\left(\boldsymbol{r}\right)\right|^{2}$
and the sum extends over the occupied states. The on-site atomic density
matrices and the smooth density are the key components in PAW and
together with the atomic information govern the updated quantities
in each SCF cycle.

In many applications, however, it is desirable to work with an orthonormal
collection of wavefunctions. As mentioned in the introduction, this
can be achieved by the transformation: 
\begin{equation}
\bar{\psi}_{m}=\hat{S}^{1/2}\tilde{\psi}_{m}\label{eq:Psibar_ShalfPsitilde}
\end{equation}
resulting in 
\begin{equation}
\bar{H}\bar{\psi}_{m}=\epsilon_{m}\bar{\psi}_{m},\label{eq:HPsibar_ePsibar}
\end{equation}
where $\bar{H}=\hat{S}^{-\frac{1}{2}}\tilde{H}\hat{S}^{-\frac{1}{2}}$.

As an example, in Fig. \ref{fig: NaCl} we show 3D isosurfaces of
$\psi$, $\bar{\psi}$ and $\tilde{\psi}$ for the $2p_{z}$ orbital
from a calculation of a single oxygen atom, as well as the associated
1D radial part obtained by projecting the 3D orbital to 1D. The three
orbitals differ only in the core region; $\psi$ clearly has more
structure in the core, while the oscillatory features are attenuated
or absent in $\bar{\psi}$ and $\tilde{\psi}$. Furthermore, the magnitude
of $\bar{\psi}$ and $\tilde{\psi}$ are smaller than that of $\psi$.

\begin{figure}
\begin{centering}
\includegraphics[width=8cm]{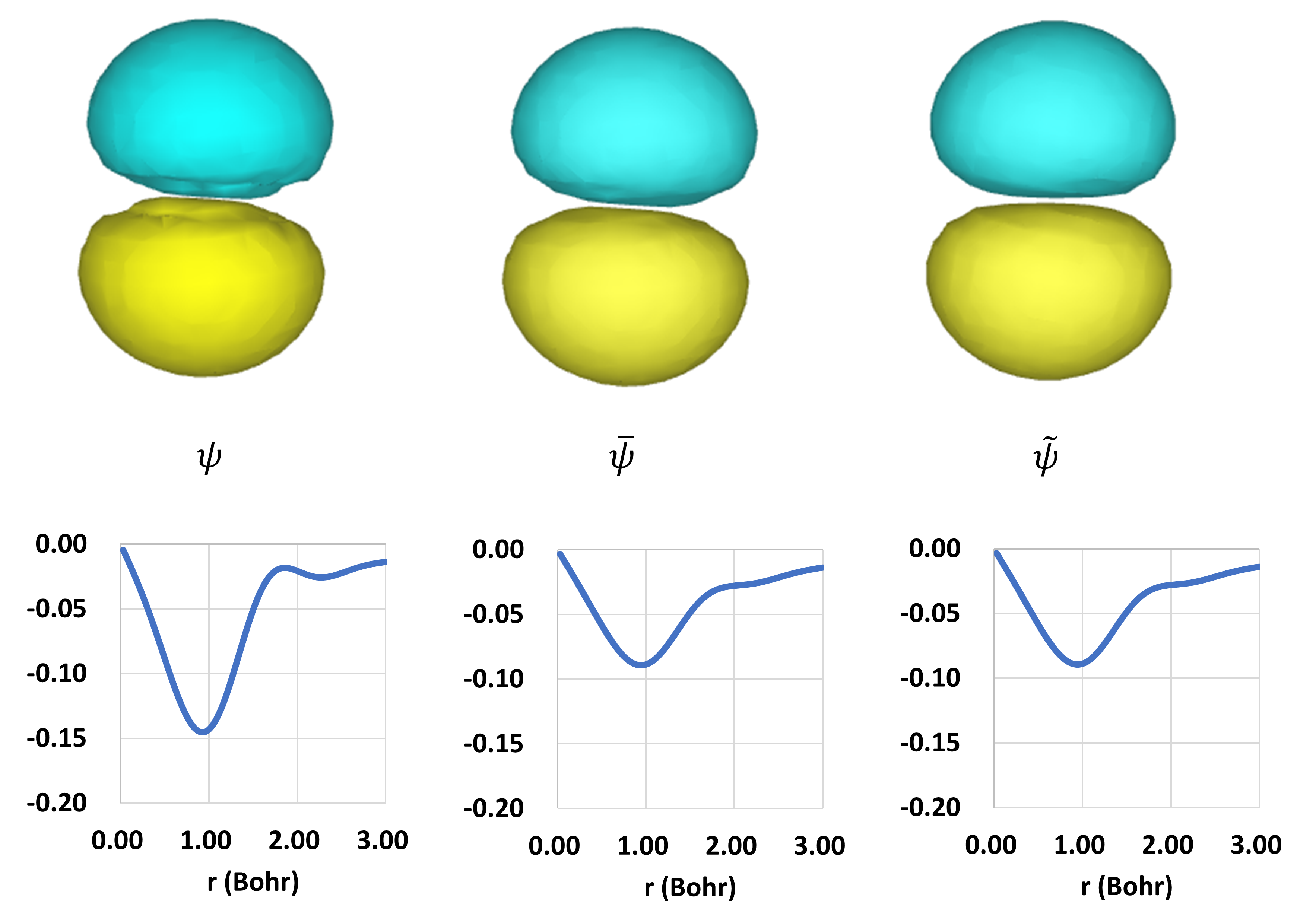}
\par\end{centering}
\caption{\label{fig:psi}Isosurfaces (top) and radial parts (bottom) of $\psi$,
$\bar{\psi}$ and $\tilde{\psi}$ for the $2p_{z}$ orbital of a single
oxygen atom. In the isosurface plot, blue color indicates positive
value, and yellow indicates negative value.}
\end{figure}

\subsubsection*{Obtaining $\hat{S}^{-1/2}$}

An efficient implementation of OPAW thus requires fast application
of $\hat{S}^{-1/2}$. For simplicity, we first consider the case where
the augmentation spheres from different atoms do not overlap, so:
$\langle p_{i}^{(a)}|p_{j}^{(a')}\rangle=0$ if $a\ne a'.$ Therefore,
we can separately rotate the $\{p_{i}^{(a)}$\} projectors around
each atoms, so that $\hat{S}$ is transformed into: 
\begin{equation}
\hat{S}=\mathbb{I}+\sum_{i,a}|\eta_{i}^{(a)}\rangle o_{i}^{(a)}\langle\eta_{i}^{(a)}|,\label{eq:Soperator}
\end{equation}
where the rotated projectors $\{\eta_{i}^{(a)}$\} are orthogonal
and satisfy $\langle\eta_{i}^{(a)}|\eta_{j}^{(a')}\rangle=\delta_{ij}\delta_{a,a'}$
(see Appendix A). With this transformation, any power of $\hat{S}$
is easily expressed; e.g., 

\begin{equation}
\hat{S}^{-\frac{1}{2}}=\mathbb{I}+\sum_{j,a}|\eta_{j}^{(a)}\rangle\left(\left(1+o_{j}^{(a)}\right)^{-\frac{1}{2}}-1\right)\langle\eta_{j}^{(a)}|.\label{eq:Smhalf}
\end{equation}
Since each $|\eta_{j}^{(a)}\rangle\langle\eta_{j}^{(a)}|$ is a projection
operator (and all such operators are orthogonal) the proof of Eq.
(\ref{eq:Smhalf}) becomes a trivial QM exercise emanating from the
simple equation $(\mathbb{I}+(a-1)P)^{m}=\mathbb{I}+(a^{m}-1)P$ when
$P$ is a projection opeator.

Next, note that the transformation operator between the orthogonal
smooth molecular orbitals and the true ones is unitary 
\begin{equation}
|\psi_{i}^{(a)}\rangle=\hat{U}|\bar{\psi}_{i}^{(a)}\rangle,\,\,\,\,\hat{U}=\hat{T}\hat{S}^{-\frac{1}{2}},
\end{equation}
so $\hat{U}^{\dagger}\hat{U}=\mathbb{I}$. Due to the unitarity, the
norm of the true molecular orbitals and the orthogonal smooth ones
is identical, as mentioned.

Overall, we note that except for the automatic orthogonality, the
algorithm is identical to the usual PAW. I.e., in an SCF cycle, with
a given one-body Hamiltonian the orthogonal molecular orbitals (the
solutions of Eq. (\ref{eq:HPsibar_ePsibar})) are first found; then,
we transform to the non-orthogonal orbitals, $\tilde{\psi}_{i}=\hat{S}^{-1/2}\bar{\psi}_{i}$
using Eq. (\ref{eq:Smhalf}), and use the usual prescription of the
PAW algorithm to update $v_{KS}(\boldsymbol{r}),D_{ij}^{(a)}$ in
the PAW Hamiltonian.

Finally, note that the assumption of non-overlapping augmentation
spheres is quite accurate, as shown in a latter section by the agreement
between our results and ABINIT. Nevertheless, it is not exact; we
could go beyond it by viewing our expression for $\hat{S}^{-\frac{1}{2}}$
as a pre-conditioner, as shown in Appendix B, and this would be pursured
in further publications.

\subsubsection*{Avoiding singularities}

The one caveat in Eq. (\ref{eq:Smhalf}) is the formal singularity
when any of the $o_{i}^{(a)}$ is close to or below $-1$. Fundamentally,
a value of $o_{i}^{(a)}=-1$ indicates that the $\hat{S}$ operator
projects out the subspace spanned by $|\eta_{i}^{(a)}\rangle o_{i}^{(a)}\langle\eta_{i}^{(a)}|$.

For a start, note that negative values of $o_{i}^{(a)}$ between -1
and 0 do not pose mathematical difficulties in our formulation, but
could indicate problems in the construction of the PAW parameters
and in the eventual implementation, depending on the PAW code used
(although they work fine in the ABINIT code used by us); see Ref.
\citet{holzwarth2019updated} for details.

In practice, for most atoms we tested, $o_{i}^{(a)}$ were well above
$-1$. We did encounter one case where $o_{i}$ is very close to $-1$
-- the GGA PAW parametrization of silicon taken from the website
of the ABINIT PAW code,\citet{jollet2014generation}\footnote{\textit{https://www.abinit.org/ATOMICDATA/014-si/Si.LDA\_PW-JTH.xml}}
where $o_{1}^{({\rm Si)}}=-1.005$. Fortunately the problem is trivially
circumvented by replacing $o_{1}^{(a)}$ by ${\rm max}(o_{1}^{(a)},-1+\delta)$
where $\delta$ is a small positive number. The results are insensitive
to $\delta$. For example, for ${\rm SiO_{2}}$ we tested (see Table
\ref{tab:delta}) three different choices, $\delta=0.003,0.01$ and
$0.05$. The two lower values of $\delta$ gave results that agree
completely with those using the LDA PAW file taken from the ABINIT
website,\citet{jollet2014generation}\footnote{\textit{https://www.abinit.org/ATOMICDATA/014-si/Si.GGA\_PBE-JTH.xml}}
where $o_{1}$ was higher than $-1$. Even the large shift parameter,
$\delta=0.05$, led to only a slight deviation.

We also note that numerical problems could also arise from the compensation
charge being negative. A solution to this problem is discussed in
the literature.\citet{holzwarth2001projector,holzwarth2019updated}

\begin{table}
\begin{centering}
\begin{tabular}{|c|c|c|c|c|c|}
\hline 
Grid spacing (Bohr) &  & 0.34 & 0.37 & 0.40 & 0.46\tabularnewline
\hline 
Gap (eV), LDA PAW &  & 5.97 & 5.97 & 5.94 & 5.85\tabularnewline
\hline 
\multirow{3}{*}{\text{Gap (eV), GGA PAW}} & $\delta=0.003$ & 5.97 & 5.97 & 5.94 & 5.85\tabularnewline
\cline{2-6} \cline{3-6} \cline{4-6} \cline{5-6} \cline{6-6} 
 & $\delta=0.01$ & 5.97 & 5.97 & 5.94 & 5.85\tabularnewline
\cline{2-6} \cline{3-6} \cline{4-6} \cline{5-6} \cline{6-6} 
 & $\delta=0.05$ & 5.95 & 5.95 & 5.92 & 5.83\tabularnewline
\hline 
\end{tabular}
\par\end{centering}
\caption{\label{tab:delta}Calculated band gaps of ${\rm SiO_{2}}$ at different
grid spacings. The Si atom PAW wavefuncton input data set based on
GGA calculations has originally $o_{1}=-1.005,$ which was modified
to $o_{1}=-1+\delta;$ different choices of $\delta$ give essentially
the same results (or slightly different for the largest $\delta$)
as does an analogous input file built based on LDA calculations where
$o_{1}>-1$. Note of course that with both data sets we did the same
overall GGA (i.e., PBE) calculation; the difference was only in the
PAW input functions}
\end{table}

\subsection{Application of OPAW in DFT and technical details}

The OPAW algorithm is general, and can be applied with any technique
requiring an orthogonal Hamiltonian. Before talking about implementation
of OPAW in DFT, note that a real space implementation of OPAW will
require the inner product between atomic projectors and wavefunctions:
$\langle p_{i}^{(a)}|\bar{\psi}\rangle$. Such inner products are
involved in determining the density matrices $\rho_{ij}^{(a)}$, as
well as applying the operators $\tilde{H}$ and $\hat{S}$. In a real
space formalism, the smooth wavefunctions $\bar{\psi}$ are defined
on a 3D grid. For computational efficiency, as long as the accuracy
of the results is not affected the grid spacing for $\bar{\psi}$
should be made as large as possible. On the other hand, the projector
functions are short-ranged and in general show larger variation than
the wavefunctions, so that evaluating the inner product directly on
a coarse 3D grid would lead to large numerical errors.

To solve this problem, we adopted the method of Ono and Hirose,\citet{ono1999timesaving}
which connects the grid of the system with a set of finer grid points
around each atom. Technical details regarding the Ono-Hirose method
are given in Appendix C.

With a real-space implementation of OPAW in hand, we applied it along
with the Chebyshev-filtered subspace iteration (CheFS) technique,\citet{zhou2014chebyshev-filtered}
resulting in an efficient DFT program (OPAW-DFT). The idea of CheFS
is described in Appendix D, along with a summary of the algorighm
in Appendix E.

Furthermore, since we are working with periodic systems, we did k-point
sampling. A brief account of using k-point sampling with OPAW is supplied
in Appendix F.

\section{Results and discussion}

\subsection{Computational details}

We did a set of calculations for periodic solids and report the calculated
fundamental band gap. The geometries are taken from the ICSD database.\footnote{\textit{https://icsd.fiz-karlsruhe.de/}}
A $4\times4\times4$ k-point mesh was used for each system.

We used the PBE GGA functional in all calculations.

For all calculations, the cutoff energy for the plane wave basis set,
$E_{{\rm cutoff}}$ is related to the density cutoff-energy by $E_{{\rm cutoff}}^{{\rm density}}=4E_{{\rm cutoff}}$,
as is typical in plane-wave calculations. Note that the latter is
related to the grid spacing for the density by $E_{{\rm cutoff}}^{{\rm density}}=\frac{1}{2}\left(\frac{\pi}{dx}\right)^{2}$.
Thus, as usual, the grid used for the density is twice as dense (in
each direction) then the spatial-grid for the plane waves.

As mentioned, to assist the SCF convergence we applied a DIIS procedure\citet{pulay1980convergence,pulay1982improved}
when updating $\nu_{KS}(\boldsymbol{r})$. At times, we have also
applied a DIIS procedure for the Hamiltonian $D_{ij}$ terms to assist
SCF convergence.

For PAW calculations, we used the recommended atomic datasets from
the ABINIT website.\citet{jollet2014generation} There are two exceptions:
the Sc atom, where the $D_{ij}$ terms were large, more than 40 Hartree,
and the Sr atom, where the $D_{ij}$ terms exceed 1000 Hartree. In
both cases this is due to a mismatch of the shape of the smooth and
true atomic orbitals in the second, outer, d-shell. To simplify, we
therefore generated new PAW potentials for Sc and Sr from the AtomPAW
package,\citet{holzwarth2001projector} using only one d-shell. For
NCPP calculations, we used the recommended pseudo-potentials from
the ABINIT website\footnote{\textit{https://www.abinit.org/psps\_abinit}}.
More information on the PAW and NCPP datasets can be found in Supplementary
Materials\footnote{See Supplementary Material at {[}URL of Supplementary Material{]}}.

\subsection{Results}

Overall, DFT calculations produce two types of information. The first
is forces and total energy, important for binding and molecular dynamics.
Here, we concentrate on the second type of output from DFT: orbital
energies and states, and here specifically the DFT HOMO-LUMO gap.
The DFT gap often serves as preliminary approximation to the actual
fundamental band gap,\citet{zhan2003ionization} and the Kohn-Sham
orbitals and their energies are the basic ingredients for most beyond-DFT
methods. Future papers will also examine the total energy and forces
with OPAW, as well as the shape of the band structure.

\begin{figure}
\begin{centering}
\includegraphics[width=8cm]{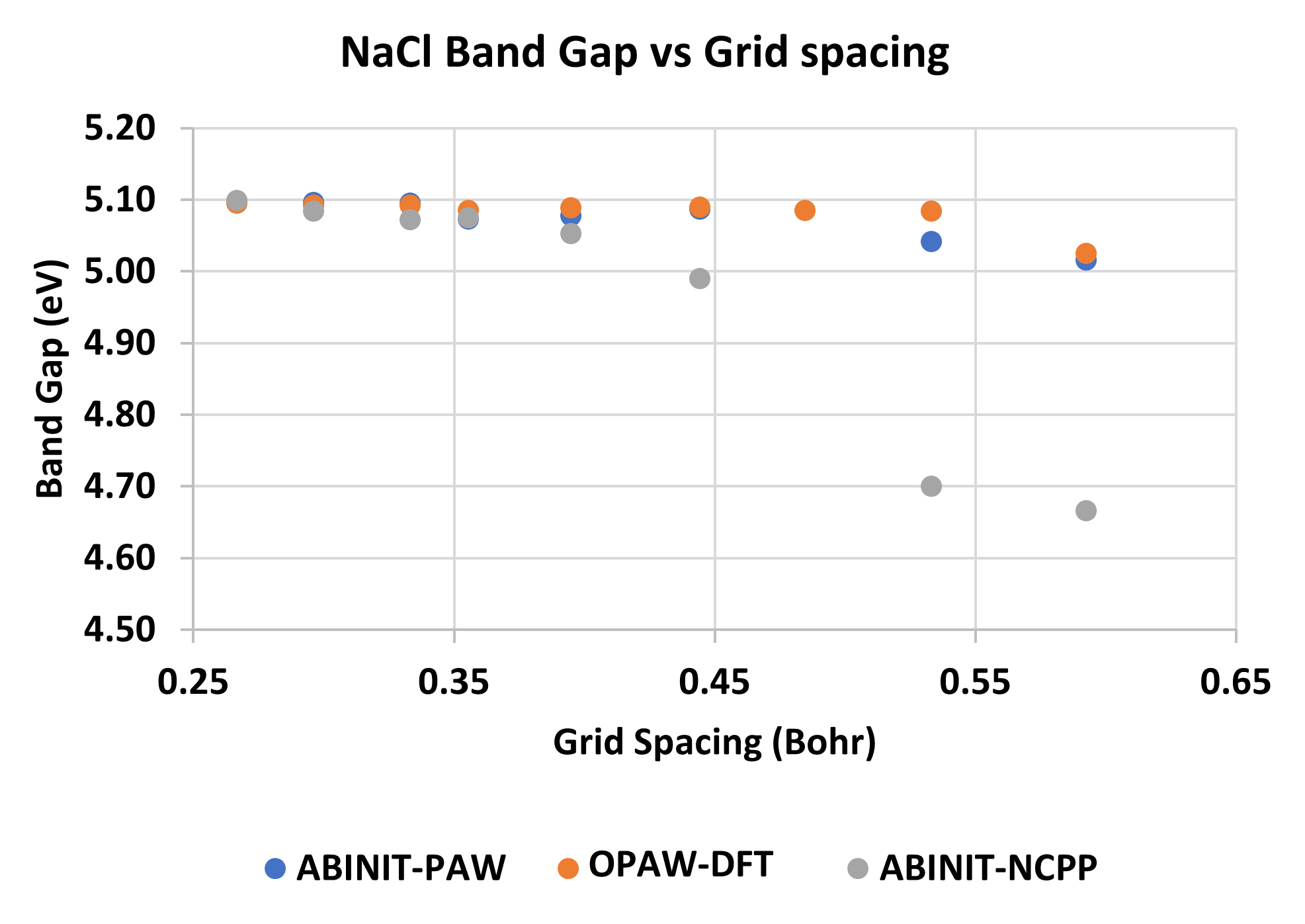}
\par\end{centering}
\caption{\label{fig: NaCl}Band gap vs. energy cutoff for NaCl, with three
methods: OPAW-DFT, ABINIT-PAW, and ABINIT-NCPP. For all the shown
cutoff energies, except the lowest one, the OPAW-DFT and ABINIT-PAW
results completely overlap on the scale of this graph.}
\end{figure}

We first examine the band-gap convergence with energy cutoff for an
NaCl solid. We compared OPAW-DFT with ABINIT simulations using PAW
or NCPP. The results are shown in Figure \ref{fig: NaCl}. For NaCl,
our OPAW-DFT successfully reproduced the ABINIT results. Furthermore,
the two PAW-based methods show better convergence with grid spacing
than the NCPP-based method.

Secondly, we report the calculated fundamental band gap of a series
of solids. A comparison of the converged results from ABINIT-PAW and
OPAW-DFT is shown in Table \ref{tab:gap-solid}. We also present the
reference value from the work of Borlido et al.\citet{borlido2019large}
The results indicate that OPAW-DFT reproduces ABINIT-PAW for a wide
variety of systems, using generally the energy cutoff in ABINIT (with
the advantage that in real-space we use the localization of the projector
functions, so the cost of appying the Hamiltonian on a single function
scales linearly with the size of the system.)

The table shows that for most solids both OPAW and ABINIT-PAW outperform
NCPP, sometimes dramatically; e.g., for SiO$_{2}$, the energy cutoff
required for converging the band gap is 15 Hartree for the two PAW
based methods, and 29 Hartree for ABINIT-NCPP calculation; for InP
te difference is even more dramatic.

\begin{table}
\begin{centering}
\begin{tabular}{|>{\centering}p{1.1cm}|>{\centering}p{1cm}|c|>{\centering}p{1cm}|>{\centering}p{1cm}|>{\centering}p{1cm}|>{\centering}p{1cm}|>{\centering}p{1cm}|}
\hline 
 & \multicolumn{2}{>{\centering}p{2cm}|}{OPAW-DFT} & \multicolumn{2}{>{\centering}p{2cm}|}{ABINIT-PAW} & \multicolumn{2}{>{\centering}p{2cm}|}{ABINIT-NCPP} & Refe-rence\citet{borlido2019large}\tabularnewline
\hline 
System & Gap & $E_{cut}$ & Gap & $E_{cut}$ & Gap & $E_{cut}$ & Gap\tabularnewline
\hline 
\hline 
NaCl & 5.09 & 11 & 5.10 & 11 & 5.07 & 25 & 5.10\tabularnewline
\hline 
CaO & 3.65 & 13 & 3.64 & 13 & 3.66 & 19 & 3.63\tabularnewline
\hline 
PbS & 0.31 & 9 & 0.29 & 9 & 0.34 & 16 & 0.30\tabularnewline
\hline 
InP & 0.68 & 10 & 0.65 & 10 & 0.69 & 23 & 0.71\tabularnewline
\hline 
Si & 0.63 & 7 & 0.63 & 7 & 0.61 & 7 & 0.62\tabularnewline
\hline 
SiO$_{2}$ & 5.99 & 15 & 5.97 & 15 & 6.00 & 29 & 6.02\tabularnewline
\hline 
ScNiSb & 0.28 & 17 & 0.25 & 15 & 0.29 & 34 & 0.30\tabularnewline
\hline 
NiScY & 0.31 & 14 & 0.28 & 14 & 0.31 & 20 & 0.30\tabularnewline
\hline 
LiH & 2.97 & 10 & 2.97 & 12 & 2.99 & 19 & 3.00\tabularnewline
\hline 
KBr & 4.33 & 8 & 4.33 & 7 & 4.34 & 18 & 4.36\tabularnewline
\hline 
K$_{3}$Sb & 0.75 & 8 & 0.74 & 5 & 0.75 & 6 & 0.77\tabularnewline
\hline 
CaCl$_{2}$ & 5.41 & 10 & 5.42 & 13 & 5.40 & 20 & 5.43\tabularnewline
\hline 
BN & 4.46 & 18 & 4.45 & 24 & 4.53 & 34 & 4.45\tabularnewline
\hline 
BaCl$_{2}$ & 5.04 & 8 & 5.04 & 8 & 5.05 & 10 & 5.03\tabularnewline
\hline 
Ar & 8.70 & 9 & 8.69 & 11 & 8.70 & 10 & 8.71\tabularnewline
\hline 
AlP & 1.58 & 9 & 1.57 & 9 & 1.58 & 12 & 1.58\tabularnewline
\hline 
SrO & 3.30 & 13 & 3.30 & 13 & 3.32 & 13 & 3.26\tabularnewline
\hline 
\end{tabular}
\par\end{centering}
\caption{\label{tab:gap-solid}Calculated fundamental band gaps (in eV) of
selected solids. The values are reported along with the planewave
cutoff (in Hartree) required for for a 0.05eV gap convergence. The
reference calculations use PAW in VASP.\citet{borlido2019large}}
\end{table}

\begin{figure}
\begin{centering}
\includegraphics[width=6cm]{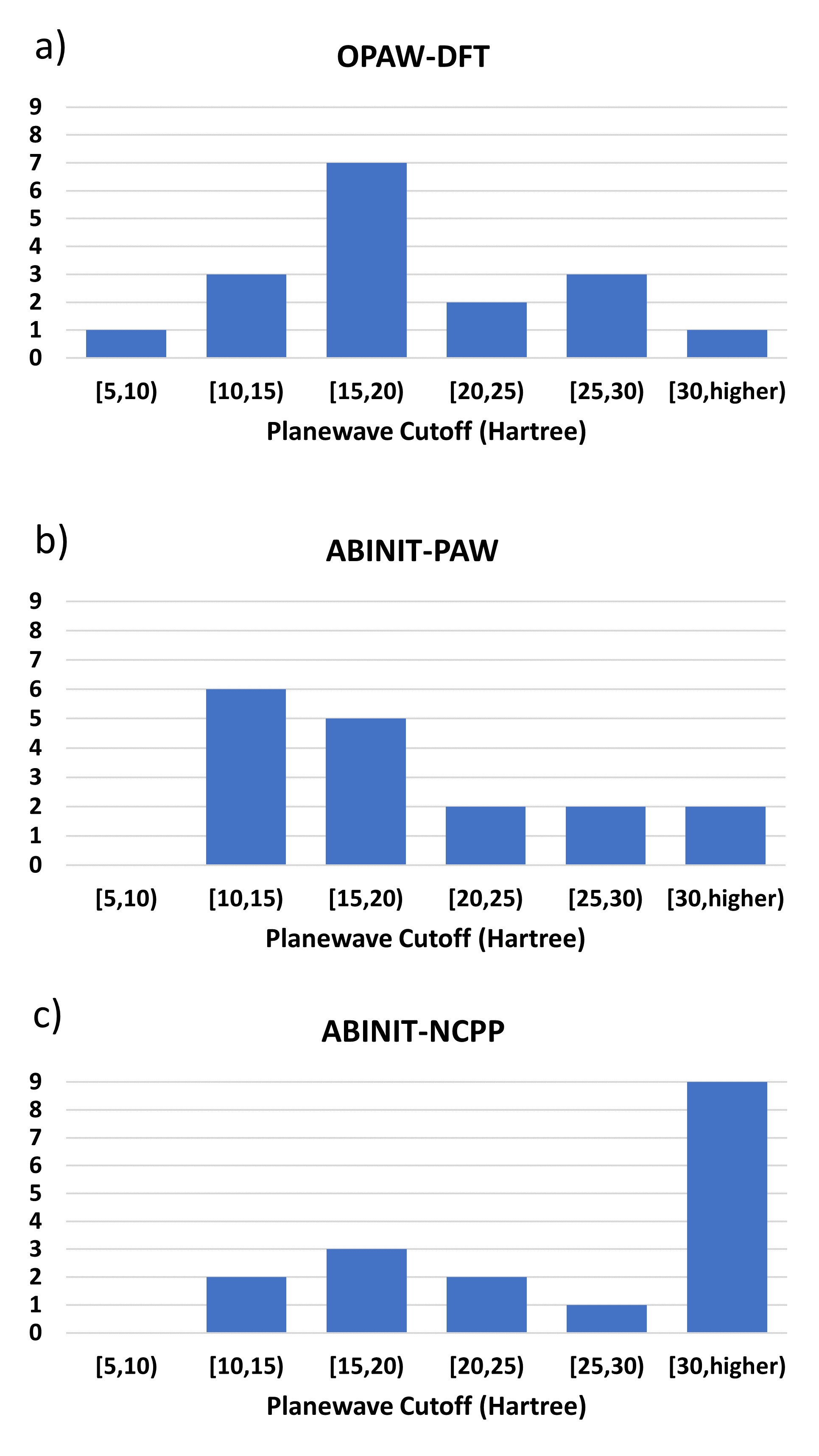}
\par\end{centering}
\caption{\label{grid spacing}Histogram of converged planewave cutoff for the
solids in Table \ref{tab:gap-solid}, from a) ABINIT-PAW; b) OPAW-DFT;
and c) ABINIT-NCPP calculations}
\end{figure}

To visualize the improvement in cutoff energy required for converging
the fundamental band gap of solids to less than $0.05$ eV, we use
histograms in Figure \ref{grid spacing}. The figure shows that PAW
gives excellent results with cutoff energies that can be as low as
7 Hartree, and are generally (in the examples we studied) below 20
Hartree.

Finally, we note that in some approaches, for example stochastic methods
for DFT, TDDFT, GW and Bethe-Salpeter \citet{baer2013self,EmbeddedFragments2014,gao2015sublinear,neuhauser2014breaking,vlcek2017stochastic,zhang2020linear,rabani2015time},
the numerical cost is related directly to the number of spatial grid
points rather than the number of plane waves; in those cases a choice
of $E_{{\rm cutoff}}^{{\rm density}}=E_{{\rm cutoff}}$ (rather than
$4E_{{\rm cutoff}}$) is better. Analog of Table \ref{tab:gap-solid}
and Figure \ref{grid spacing} for this choice can be found in the
Supplementary Material \footnote{See Supplementary Material at{[}URL of Supplementary Material{]}}.
On average the $E_{{\rm cutoff}}^{{\rm density}}$ required when $E_{{\rm cutoff}}^{{\rm density}}=E_{{\rm cutoff}}$
is much smaller than that required when using $E_{{\rm cutoff}}^{{\rm density}}=4E_{{\rm cutoff}}$
(as done above), i.e. setting $E_{{\rm cutoff}}^{{\rm density}}=E_{{\rm cutoff}}$
allows a much sparser real space grid.

\section{Conclusions}

The results in the previous section show that our efficient OPAW reproduces
traditional PAW. The OPAW algorithm is easy to implement and combines
the best of both worlds: the lower cutoff energy typically enabled
by PAW and the orthogonality of norm-conserving pseudopotential approaches.

With the efficient methodology for acting with the Hamiltonian and
overlap/inverse overlap, i.e., the simple application (on any function
$f$) of $\hat{S}f$, $\hat{H}f$ $\hat{S}^{-1}f$, $\hat{S}^{-\frac{1}{2}}f$
and $\hat{S}^{-\frac{1}{2}}\tilde{H}\hat{S}^{-\frac{1}{2}}f$, we
can combine PAW with other electronic structure theory methods, including
our linear scaling stochastic TDDFT and GW methods,\citet{neuhauser2014breaking,vlcek2017stochastic,gao2015sublinear}
opening the door to significant (in some cases an order of magnitude)
improvements in overall grid size and the reduction of the spectral
range, and potentially even larger improvements in the cost of beyond-DFT
approaches.

Finally, we note that an example where some of the developments here
were applied is our recent large scale stochastic long-range exchange
method for TDDFT using PAW.\citet{zhang2020linear}

\section*{Acknowledgements}

We are grateful to Roi Baer, Eran Rabani, Vojtech Vlcek and Xu Zhang
for helpful conversations. This work was supported by the NSF CHE-1763176
grant. Computational resources were supplied through the XSEDE allocation
TG-CHE170058.

\section*{Appendix A: Transformation through $\hat{S}$}

\renewcommand{\theequation}{A.\arabic{equation}}
\setcounter{equation}{0}

We start by a proof of Eq. (\ref{eq:psitilde_S_psitilde}). Since
the molecular orbitals are orthogonal, $\left\langle \psi_{i}|\psi_{j}\right\rangle =\delta_{ij},$
and since $|\psi_{i}\rangle=\hat{T}|\tilde{\psi}_{i}\rangle$, it
follows that $\left\langle \tilde{\psi}_{i}|\hat{T}^{\dagger}\hat{T}|\tilde{\psi}_{i}\right\rangle =\delta_{ij},$
which given the definiton $\hat{S}\equiv\hat{T}^{\dagger}T$ yields
Eq. (\ref{eq:psitilde_S_psitilde}).

In the remainder we discuss the technical details of the transformation.

Given the initial operator: 
\begin{equation}
\hat{S}=\mathbb{I}+\sum_{ij,a}|p_{i}^{(a)}\rangle s_{ij}\langle p_{j}^{(a)}|,\label{eq:A1}
\end{equation}
the first step is to orthonormalize the projectors. For each atom,
define a projector overlap matrix $L_{ij}^{(a)}=\langle p_{i}^{(a)}|p_{j}^{(a)}\rangle$,
and diagonalize it: $L^{(a)}=U^{(a)}\lambda^{(a)}U^{(a)\dagger}$,
with $U^{(a)}$ unitary. Then, define a new set of projectors $\{\xi_{i}^{(a)}\}$:
\begin{equation}
|\xi_{i}^{(a)}\rangle=\frac{1}{\sqrt{\lambda_{i}^{(a)}}}\sum_{j}U_{ji}^{(a)}|p_{j}^{(a)}\rangle\label{eq:A2}
\end{equation}
that will be orthogonal, $\langle\xi_{i}^{(a)}|\xi_{j}^{(a)}\rangle=\delta_{ij}.$
Inverting Eq. (\ref{eq:A2}) and substituting into Eq. (\ref{eq:A1})
then gives: 
\begin{equation}
\hat{S}=\mathbb{I}+\sum_{kl,a}|\xi_{k}^{(a)}\rangle O_{kl}^{(a)}\langle\xi_{l}^{(a)}|\label{eq:A3}
\end{equation}
where $O^{(a)}=\sqrt{\lambda^{(a)}}U^{(a)}s^{(a)}U^{(a)\dagger}\sqrt{\lambda^{(a)}}$.

The next step involves diagonalization of the matrix $O^{(a)}$, as
$O^{(a)}=Q^{(a)}o^{(a)}Q^{(a)\dagger}$, with $Q^{(a)}$ unitary.
It then readily follows that: 
\begin{equation}
\hat{S}=\mathbb{I}+\sum_{i,a}|\eta_{i}^{(a)}\rangle o_{i}^{(a)}\langle\eta_{i}^{(a)}|,\label{eq:A4}
\end{equation}
where $|\eta_{i}^{(a)}\rangle=\sum_{l}Q_{li}^{(a)}|\xi_{l}^{(a)}\rangle$
are also orthogonal due to the unitarity of $Q^{(a)}.$ (Note that
a diagonal representation of projectors is also done in NCPP, where
diagonal projectors are used in representing the non-local potential.\citet{hamann2013optimized})

Finally, when we apply the Ono-Hirose procedure, the bare $\eta_{i}^{(a)}$
are replaced by the processed ones, $\bar{\eta}_{i}^{(a)}$ as in
Eq. (\ref{eq:p_fromfine}), i.e., 
\begin{equation}
\hat{S}=\mathbb{I}+\sum_{i,a}|\bar{\eta}_{i}^{(a)}\rangle o_{i}^{(a)}\langle\bar{\eta}_{i}^{(a)}|.\label{eq:A5}
\end{equation}
These are not orthogonal on the rough-grid surrounding each molecule.
We therefore repeat the orthogonalization procedure, Eqs. (\ref{eq:A1})-(\ref{eq:A4}),
with the overlap matrix $L^{(a)}$ now being replaced by $\bar{L}_{ij}^{(a)}=dv\sum_{\boldsymbol{r}}\bar{\eta}_{i}^{(a)}(\boldsymbol{r)}\bar{\eta}_{j}^{(a)}(\boldsymbol{r)}$,
leading eventually to 
\begin{equation}
\hat{S}=\mathbb{I}+\sum_{i,a}|\bar{\zeta}_{i}^{(a)}\rangle\bar{o}_{i}^{(a)}\langle\bar{\zeta}_{i}^{(a)}|,\label{eq:A6}
\end{equation}
where $\bar{\zeta}_{i}^{(a)}$ are orthogonal on the rough grid, $\langle\bar{\zeta}_{i}^{(a)}|\bar{\zeta}_{j}^{(a)}\rangle=\delta_{ij}$.

\section*{Appendix B: Going beyond the non-overlapping augmentation spheres
assumption}

\renewcommand{\theequation}{B.\arabic{equation}}
\setcounter{equation}{0}

In this appendix we show how one could go beyond the non-overlapping
augmentation sphere assumption. Let's consider for simplicity exprssions
using $\hat{S}^{-1}$ rather than $\hat{S}^{-\frac{1}{2}}$. Then,
the generic relation $\hat{S}\psi=H\xi$ (the inversion of which is
the crucial step in a Chebyshev propagation that iterates $\hat{S}^{-1}H$
) can be rewritten as

\begin{equation}
(\mathbb{I}+\hat{B})\psi=\xi'\label{eq:Sinv_preconditioner}
\end{equation}
where $\xi'\equiv\hat{S}_{NO}^{-1}H\xi,$ and

\begin{equation}
\hat{B}\equiv\hat{S}_{NO}^{-1}\hat{S}-\mathbb{I},
\end{equation}
while $\hat{S}_{NO}^{-1}$ is a non-ovelapping ($NO$) expression
for $\hat{S}^{-1},$ as in Section II.A

\begin{equation}
\hat{S}_{NO}^{-1}=\mathbb{I}+\sum_{J}\left((o_{J}+1)^{-1}-1\right)P_{J},\label{eq:SNOinv}
\end{equation}
and we use the abbreviated notation from there (but without assuming
that different $P_{J}$ are orthorgonal). Note that this appendix
is the only place in the paper where we give an explicit subscipt
($NO$) to expressions obtained under the non-overlapping assumption.

Equation (\ref{eq:Sinv_preconditioner}) could be solved by a Taylor
expression in $B,$which measures the deviation from the non-overlapping
spheres assumption. Recall that our results, obtained essetnially
by assuming that $B=0$, are all quite accurate. Therefore, even a
single Taylor term should be extremely accurate, i.e.,

\begin{equation}
\psi=(\mathbb{I}-\hat{B})\xi'=(2\mathbb{I}-\hat{S}_{NO}^{-1}\hat{S})\hat{S}_{NO}^{-1}H\xi,
\end{equation}
and as a reminder the definitons of the terms here come from Eqs.
(\ref{eq:sij}),(\ref{eq:H}) and (\ref{eq:SNOinv}). This expression
would not be much more expensive than the $B=0$ expression we used
throughout the rest of the paper ($\psi=\hat{S}_{NO}^{-1}H\xi$),
since it only differs in the use of further overlaps.

\section*{Appendix C: The Ono-Hirose transformation with a spline method and
its implications in OPAW}

\renewcommand{\theequation}{C.\arabic{equation}}
\setcounter{equation}{0}

The method of Ono and Hirose\citet{ono1999timesaving} is used to
connect, for each atom, two sets of local grids. (The grids are specific
to each atom, but for brevity we omit the atomic label in the following
derivations.) One is a 'rough grid' $X^{r}$, consisting of a small
cubic region of the 3D wavefunction grid, which encloses the augmentation
sphere for the specific atom. The second is a 'fine grid' $X^{f}$,
spanning the same volume but with more grid points and smaller grid
spacing.

The overlap of the waveunctions and projectors should formally be
performed on the fine grid. This requires, formally, interpolating
the wavefunction from the rough grid (i.e., $\psi(\boldsymbol{r}),\boldsymbol{r}\in X^{r}$)
to the fine grid, as 
\begin{equation}
\psi(\boldsymbol{r}_{f})=\sum_{\boldsymbol{r}\in X^{r}}B\left(\boldsymbol{r}_{f},\boldsymbol{r}\right)\psi(\boldsymbol{r}),\label{eq:OnoHirose}
\end{equation}
where $B\left(\boldsymbol{r}_{f},\boldsymbol{r}\right)$ is a linear
projection matrix. Earlier applications of the Ono-Hirose approach
usually used cubic fitting,\citet{ono1999timesaving,enkovaara2010electronic,mortensen2005real}
but here we used a spline fit.

The key observation of the Ono-Hinose approach is then that the fine-grid
overlap of the atomic projectors and the wavefunctions, 
\[
\langle p_{i}^{(a)}|\psi\rangle\equiv\sum_{\boldsymbol{r}_{f}\in X^{f}}p_{i}^{(a)}(\boldsymbol{r}_{f})\psi(\boldsymbol{r}_{f})dv_{f},
\]
can be written as a rough-grid overlap 
\begin{equation}
\langle p_{i}^{(a)}|\bar{\psi}\rangle=\sum_{\boldsymbol{r}\in X^{r}}\bar{p}_{i}^{(a)}(\boldsymbol{r})\bar{\psi}(\boldsymbol{r})dv,
\end{equation}
where $dv_{f}$ and $dv$ are the fine-grid and rough-grid volume
elements, and 
\begin{equation}
\bar{p}_{i}^{(a)}(\boldsymbol{r})=\frac{dv_{f}}{dv}\sum_{\boldsymbol{r}_{f}\in X^{f}}p_{i}^{(a)}(\boldsymbol{r}_{f})B(\boldsymbol{r}_{f},\boldsymbol{r}).\label{eq:p_fromfine}
\end{equation}

The key practical aspect in the Ono-Hirose transformation is the smoothing
matrix, $B(\boldsymbol{r}_{f},\boldsymbol{r})$, connecting the fine
and rough grids (Eq. (\ref{eq:OnoHirose})). Typically a cubic-fit
approach is used; here we opted instead to use a spline fit matrix,
which is separable. 
\begin{equation}
B(\boldsymbol{r}_{f},\boldsymbol{r})=\beta(x_{f},x)\beta(y_{f},y)\beta(z_{f},z),\label{eq:Bspline_separable}
\end{equation}
where the $\beta$ matrices are obtained as explained below, and depend
on the element only, not the specific atoms (the derivation is done
for the case of equal grid spacings, $dx=dy=dz$, and is trivially
extended in the general case).

For each different element a small padding region is added around
the augmented region (typically of size $r_{{\rm pad}}=$0.5 or 1Bohr,
the results do not change if either value is used). Then the set of
all $x$ points within a distance $\pm\bar{r}$ from the nucleus,
where $\bar{r}=r_{{\rm aug}}+r_{{\rm pad}}$, is labeled as $\left\{ x_{i}\right\} _{i=1,...,n_{1d}}$.
Here, $n_{1d}\simeq2\frac{\bar{r}}{dx}$, and will be typically 6-14
for our grid parameters. The set $\left\{ x_{i}\right\} _{i=1,...,n_{1d}}$
will be denoted as the rough-1d grid in the $x$ direction.

We define then a fine 1D grid of size $n_{f}=1+(n_{1d}-1)m_{f},$
where $m_{f}$ is adjusted so that the fine grid spacing, $dx_{f}=\frac{dx}{m_{f}}$
is quite small, about $0.1-0.15$Bohr (thus typically $n_{f}\sim20-50$).
Further, we relabel $\beta(x_{f},x)$ as a matrix, $\beta(i_{f},i),$with
$1\le i\le n_{1d}$, $1\le i_{f}\le n_{f}$.

The $\beta(i_{f},i)$ matrix is formally defined as the spline fit
coefficient matrix, i.e., given a 1-d function $g(x_{i})$ on a rough
grid$,$ then the fine-grid spline interpolation is 
\begin{equation}
g(x_{i_{f}})=\sum_{i}\beta(i_{f},i)g(x_{i}).
\end{equation}

While it is possible to derive $\beta(i_{f},i)$ formally, the simplest
approach is to use a set of delta-functions. For example, to obtain
$\beta(i_{f},i=1)$ use a spline fit subroutine with a $g(x_{i})=\delta_{1,i}$
input vector, feed it to a spline-fit interpolation program, and the
resulting $g(x_{i_{f}})$ fine-grid vector will be exactly $\beta(i_{f},i)$
for $i=1$.

Given the $\beta(i_{f},i)$ matrix (now again relabeled as $\beta(x_{f},x)$),
the next stage is to rotate each fine-grid function to the rough grid,
Eq. (\ref{eq:p_fromfine}). This is easily done in stages due to the
separability of Eq. (\ref{eq:Bspline_separable}), so that the total
cost to transform each function is only about $n_{f}^{3}n_{1d},$
which works out to be about a one-time cost of 3,000-100,000 operations
for each atom and for each projector, i.e., an overall negligibly
small cost.

A side note: as it stands Eq. (\ref{eq:Bspline_separable}) and therefore
the remainder of our derivation only applies to orthogonal cells;
however, it is trivially generalized to other cyrstallographic cells,
by replacing $x,y,z$ by non-orthogonal coordinates that are parrallel
to the unit cell directions.

Finally, we note that there are alternatives to the Ono-Hirose technique,
primarily the Mask Function Technique, where the radial functions
are smoothed.\citet{tafipolsky2006general}

\section*{Appendix D: Chebyshev-filtered subspace iteration}

\renewcommand{\theequation}{D.\arabic{equation}}
\setcounter{equation}{0}

The OPAW algorithm is general, and can be applied with any technique
requiring an orthogonal Hamiltonian. Here we combined our OPAW approach
with the Chebyshev-filtered subspace iteration (CheFS) technique\citet{zhou2014chebyshev-filtered}
resulting in an efficient DFT program (OPAW-DFT).

In CheFS, with each iteration a more refined subspace is obtained,
spanned by the lower energy orbitals. The Chebyshev filter 
\[
F_{J}(\bar{H})=C_{J}\left[\frac{\bar{H}-\frac{c+b}{2}\mathbb{I}}{\frac{c-b}{2}}\right]
\]
selectively enhances the occupied orbitals. Here $C_{J}$ is a Chebyshev
polynomial of degree $J$ (typically taken as $J\approx20$) and its
argument is a shifted Hamiltonian, where $b$ is set to be a little
bit higher than LUMO energy and $c$ is set to be higher than the
maximum eigenvalue of $\bar{H}$. The filter magnifies the weight
of the lower end of the spectrum (energies below $b$). The number
of states that the filter is operated on, labeled $M$, needs to be
somewhat larger than the number of occupied molecular orbitals.

Obtaining the action of $F_{J}(\bar{H})$ on a function involves repeated
applications of $\bar{H}$. In practice, we could either apply $F_{J}(\bar{H})$
directly, or note that this is equivalent to $S^{1/2}F_{J}(\hat{S}^{-1}\tilde{H})S^{-1/2}$.
The latter is numerically slightly more efficient, since it involves
only one application of an $S$-type projector; practically, to obtain
$\hat{S}^{-1}$ one simply need to replace the $-\frac{1}{2}$ powers
in Eq. (\ref{eq:Smhalf}) by $-1$. We verified that the two techniques
give numerically the same results.

A summary of the structure of the OPAW-DFT algorithm is given next.

\section*{Appendix E: Summary of algorithm}

\renewcommand{\theequation}{E.\arabic{equation}}
\setcounter{equation}{0}

For a given system, first,
\begin{itemize}
\item At this stage $a$ refers to each element in the system. From a given
data set of atomic $\phi_{i}^{(a)},\phi_{i}^{(a)},p_{i}^{(a)}$ (typically
contained in an ``XML'' file) construct the $s_{ij}^{(a)}$ matrix,
as well as several small-atom matrices needed for the PAW algorithm.
Construct a new set of orthogonal orbitals, $\eta_{i}^{(a)}$, that
are a linear combination of $p_{i}^{(a)}$, and extract the $o_{i}^{(a)}$
coefficients (Appendix A). Shift $o_{i}^{(a)}$ to be above -1 if
necessary.
\item Starting at this next stage, $a$ refers to each atom separately.
Use the Ono-Hirose transformation (Appendix C) to form $\bar{p}_{i}^{(a)}(\boldsymbol{r}),$
each on a small rough-grid around each atom. Similarly form $\bar{\eta}_{i}^{(a)},$
and orthogonalize them (Appendix C) to form $\bar{\zeta}_{i}^{(a)}(\boldsymbol{r})$
that are orthogonal on the grid. A new set of $\bar{o}_{i}^{(a)}$
is then produced; again shift each $\bar{o}_{i}^{(a)}$ to be above
-1 if necessary.
\end{itemize}
Then start the SCF algorithm, presented first in terms of the orthogonal
Hamiltonian, $\bar{H}$. All expressions now refer to the sparse 3D
grid.

Pick a set of $M$ random plane-wave orbitals, $\bar{\psi}_{m\boldsymbol{k}}(\boldsymbol{q})\boldsymbol{.}$
(See Appendix F for details of the k-point sampling.) Orthogonalize
them, and then do the following loop till convergence:
\begin{itemize}
\item Fourier transform the orbitals to the equivalent density-based spatial
grid, $\psi_{m\boldsymbol{k}}(\boldsymbol{r})$. Form $\tilde{\psi}_{m\boldsymbol{k}}(\boldsymbol{r})=\langle\boldsymbol{r}|S^{\frac{1}{2}}|\psi_{m\boldsymbol{k}}\rangle.$
\item From $\tilde{\psi}_{m\boldsymbol{k}}(\boldsymbol{r})$, calculate
the atomic density-type matrices, $\rho_{ij}^{(a)}$ and construct
the smooth density, DFT potential, and the $D_{ij}^{(a)}$ terms.
We adopted the routines of ABINIT for this stage.
\item Starting at the 2nd iteration, we apply at this stage a DIIS iteration
on the DFT potential, $v_{KS}(\boldsymbol{r}),$ and potentially also
on the $D_{ij}$ terms.
\item Apply the $J$-th degree Chebyshev operator; symbolically assign $\bar{\psi}_{m\boldsymbol{k}}\leftarrow F_{J}\text{\ensuremath{\left(\bar{H}^{\boldsymbol{k}}\right)}}\bar{\psi}_{mk}$.
This could be done either totally at the spatial grid level, $\bar{\psi}_{m\boldsymbol{k}}(\boldsymbol{r}),$
or alternately, one could at each stage (i.e., after each application
of $\bar{H}^{\boldsymbol{k}}$) transfer back to the plane-wave grid,
$\bar{\psi}_{m\boldsymbol{k}}(\boldsymbol{q})$ keeping only values
of $\boldsymbol{q}$ with energies below $E_{{\rm cutoff}}$ and then
convert back to $\bar{\psi}_{m\boldsymbol{k}}(\boldsymbol{r})$. There
is no difference in the accuracy using either approach.
\item At the end of the Chebyshev iteration, transfer to the plane-wave
grid, orthogonalize the resulting functions $\bar{\psi}_{m\boldsymbol{k}}(\boldsymbol{q})$,
rotate back to $\boldsymbol{r}$ space, diagonalize the $M\times M$
matrix $h_{mm'}^{\boldsymbol{k}}=\left\langle \bar{\psi}_{m\boldsymbol{k}}\left|\bar{H}^{\boldsymbol{k}}\right|\bar{\psi}_{m'\boldsymbol{k}}\right\rangle $
in the resulting basis of $M$ vectors, and rotate $\bar{\psi}_{m\boldsymbol{k}}(\boldsymbol{q})$
accordingly (with the resulting vectors again labeled $\bar{\psi}_{m\boldsymbol{k}}(\boldsymbol{q}))$.
\item Based on the resulting orbital energies, assign occupation numbers.
Repeat the cycle till SCF convergence (typically 10-20 times).
\end{itemize}
The algorithm is only slightly modified if we choose to replace the
orthogonal $\bar{H}^{\boldsymbol{k}}$ by $\left(\hat{S}^{\boldsymbol{k}}\right)^{-1}\tilde{H}^{\boldsymbol{k}}$.
In that case the only modifications are that we directly iterate $\tilde{\psi}_{m\boldsymbol{k}}\leftarrow F_{J}\left(\left(\hat{S}^{\boldsymbol{k}}\right)^{-1}\tilde{H}^{\boldsymbol{k}}\right)\tilde{\psi}_{m\boldsymbol{k}}$,
and at the end of each Chebyshev series we need to use general orthogonalization,
so $\left\langle \tilde{\psi}_{m\boldsymbol{k}}\left|\hat{S}^{\boldsymbol{k}}\right|\tilde{\psi}_{m'\boldsymbol{k}}\right\rangle =\delta_{mm'}.$

\section*{Appendix F: k-point sampling}

\renewcommand{\theequation}{F.\arabic{equation}}
\setcounter{equation}{0}

For periodic systems, the plane-wave wavefunctions are given by Bloch
waves, $e^{i\boldsymbol{k\cdot r}}\bar{\psi}_{m\boldsymbol{k}}(\boldsymbol{r})$
where $\boldsymbol{k}$ samples the first Brillouin zone, and $\bar{\psi}_{m\boldsymbol{k}}(\boldsymbol{r})$
are periodic. The modifications are therefore straightforward, exactly
analogous to PAW and NCPP: Given a periodic Bloch state $\bar{\psi}_{mk}(\boldsymbol{r})$
on a 3D unit cell grid, define a $\boldsymbol{k}$-dependent Hamiltonian
as $\bar{H}^{\boldsymbol{k}}=\left(\hat{S}^{\boldsymbol{k}}\right)^{-\frac{1}{2}}\tilde{H}^{\boldsymbol{k}}\left(\hat{S}^{\boldsymbol{k}}\right)^{-\frac{1}{2}}$,
with (in the spatial basis):
\begin{equation}
\left(\hat{S}^{\boldsymbol{k}}\right)^{-\frac{1}{2}}|\bar{\psi}_{m\boldsymbol{k}}\rangle=|\bar{\psi}_{m\boldsymbol{k}}\rangle+e^{-i\boldsymbol{k\cdot r}}\sum_{i,a}|\bar{\zeta}_{i}^{(a)}\rangle\bar{o}_{i}^{(a)}\langle\bar{\zeta}_{i}^{(a)}|e^{i\boldsymbol{k\cdot r}}\bar{\psi}_{m\boldsymbol{k}}\rangle.
\end{equation}
I.e., in each application the $\bar{\psi}_{m\boldsymbol{k}}$ molecular
orbital is multiplied once by $e^{i\boldsymbol{k\cdot r}}$, the projection
performed for all atoms, and the resulting orbital is multiplied again
by $e^{-i\boldsymbol{k\cdot r}}.$

Within the $\tilde{H}^{\boldsymbol{k}}$ operator, the $D_{ij}$ terms
are similarly calculated, and the kinetic energy with the kinetic
energy operator obtained as usual by passing to Fourier space (i.e.,
producing $\bar{\psi}_{j\boldsymbol{k}}(\boldsymbol{G})$), multiplying
by $\frac{1}{2}(\boldsymbol{k+G})^{2}$, and transforming back.

\bibliographystyle{apsrev4-1}\bibliographystyle{apsrev4-1}
\bibliography{ref}

\end{document}